\documentclass[aps,prl,twocolumn,showpacs,nofootinbib,superscriptaddress]{revtex4}
\usepackage{graphicx}
\usepackage{color}
\usepackage{epsfig}
\usepackage{tikz}
\usepackage{amsmath,amssymb}
\usepackage{dcolumn,bm}

\def\[{\left[}
\def\]{\right]}
\def\({\left(}
\def\){\right)}

\newcommand{\iisc}
{\affiliation{Centre for Condensed Matter Theory, Department of Physics, Indian Institute of Science, Bangalore 560012, India}}

\newcommand{\rri}
{\affiliation{Raman Research Institute, Bangalore 560080, India}}

\newcommand{\ncbs}
{\affiliation{National Centre for Biological Sciences (TIFR), Bangalore 560065, India}}

\begin{document}
\title
{Active fluidization in dense glassy systems}

\author{Rituparno Mandal}%
\email[Email: ]{rituparno@physics.iisc.ernet.in}
\iisc

\author{Pranab Jyoti Bhuyan}%
\email[Email: ]{pranab@physics.iisc.ernet.in}
\iisc

\author{Madan Rao}%
\email[Email: ]{madan@ncbs.res.in}
\rri
\ncbs

\author{Chandan Dasgupta}%
\email[Email: ]{cdgupta@physics.iisc.ernet.in}
\iisc

\begin{abstract}
Dense soft glasses show strong collective caging behavior at sufficiently 
low temperatures. Using molecular dynamics simulations of a model glass former, 
we show that the incorporation of activity or self-propulsion, $f_0$, can 
induce cage breaking and fluidization, resulting in a disappearance of the 
glassy phase beyond a critical $f_0$. The diffusion coefficient crosses over 
from being strongly to weakly temperature dependent as $f_0$ is increased. 
In addition, we demonstrate that activity induces a crossover from a fragile 
to a strong glass and a tendency for clustering of active particles. 
Our results are of direct relevance to the collective dynamics of dense active 
colloidal glasses and to recent experiments on tagged particle diffusion in 
living cells.

\end{abstract}

\pacs{61.20.Ja, 64.70.D, 64.70.Q-}
\maketitle

Disordered assemblies of particles approaching the glass transition, either 
by lowering the temperature or by increasing the density, exhibit increasingly 
slow dynamics and strong caging of tagged particle movement. Recently, there 
has been a lot of interest in the collective behavior of dense assemblies
of hard-sphere-like active or self-propelled particles close to the 
glass transition~\cite{Berthier:14,Ni:13,Marchetti:11,Kranz:10}. 
In these systems, the glass transition is approached 
on increasing the packing fraction and the introduction of activity tends 
to fluidize the assembly by shifting the glass transition to higher packing 
fractions. The corresponding behavior in thermally controlled  
systems in which the
glass transition occurs upon decreasing the temperature at constant density 
has not been explored till now.

Recent experimental studies of tagged particle diffusion in living cells make 
this question most pertinent. 
For instance, studies on tagged particle diffusion of cytoplasmic constituents 
revealed that bacterial cytoplasm exhibits characteristic glassy features 
such as caging, non-ergodicity and dynamical heterogeneity ~\cite{Bradley:14} 
in the absence of metabolic activity, and shows liquid-like features when 
subjected to activity through cellular metabolism. Likewise, microrheology 
studies of particles embedded in the cell nucleus show cage-hopping dynamics 
driven by active stress fluctuations arising from ATP-dependent chromatin 
remodeling proteins (CRPs) and complete caging in ATP-depleted conditions 
(or when the activity of CRPs is perturbed)~\cite{Rao:12}. Recent 
microrheology work on gold particles attached to cytoskeleton-motor complexes 
shows the effects of non-thermal fluctuations in transporting particles over 
large scales~\cite{Schmidt:07,weitz:14}. Such activity driven fluidization 
could play a significant role in transporting molecules over large scales, 
thereby regulating a variety of biochemical signaling reactions within 
the cell \cite{Abhishek}.

\begin{figure}
\includegraphics[height = 0.56\linewidth]{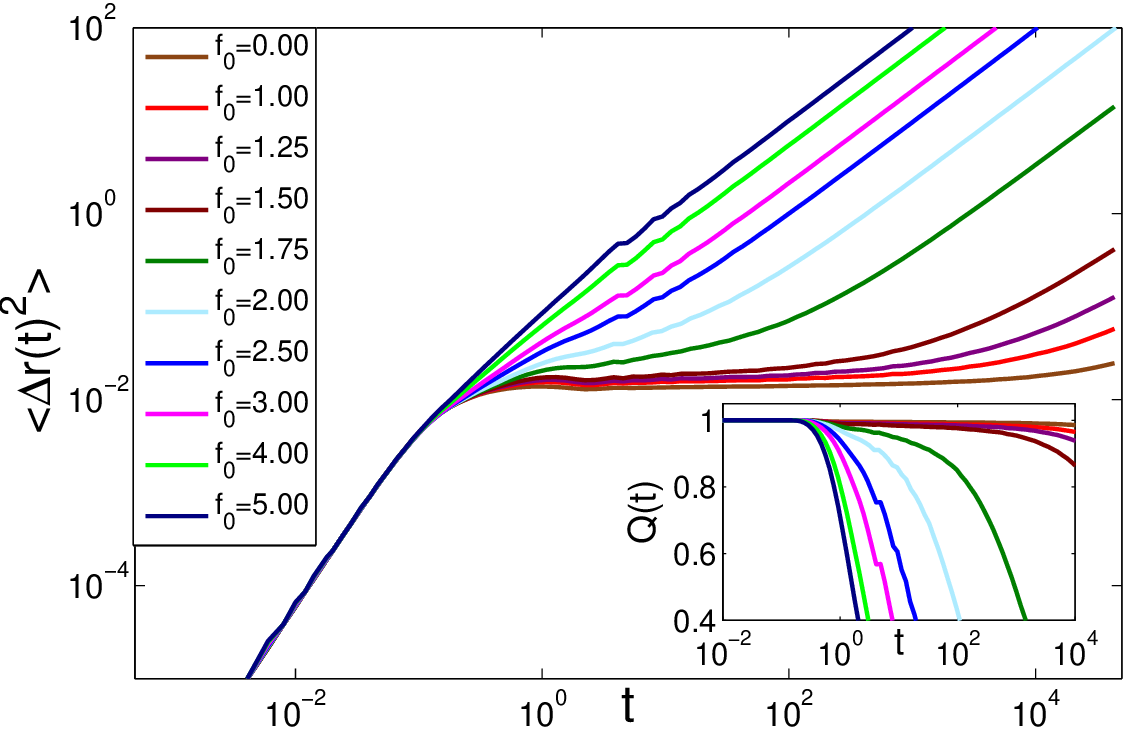}
\caption{(color online). MSD for the active Kob-Andersen model for
different values of the self-propulsion force $f_{0}$ for $T=0.2$, $\rho_a=1.0$
and $\tau_{p}=4.0$. Increasing activity induces cage escape and a crossover 
to late time diffusive behavior. Inset: Plots of the overlap function $Q(t)$ for different 
self-propulsion forces $f_{0}$ show a similar behavior.}
\label{msdqt}
\end{figure}

Given the generality of these observations, we have studied, using molecular
dynamics (MD) simulations, the effects of 
active driving in a generic model glass former, the Kob-Andersen 
binary mixture~\cite{Kob:94}. Activity is introduced by assuming that a 
fraction $\rho_a$ of one kind of particles in the binary system experience a random active 
force $f_0$ that is correlated over a persistence time $\tau_{p}$. 
Our main results are : (i) Activity fluidizes the glass and dramatically 
reduces the glass transition temperature; (ii) Tagged particle dynamics shows 
cage-hopping resulting in a late time diffusion coefficient that is weakly 
dependent on the temperature in the limit of large activity; (iii) The effect of activity on dynamical 
correlation functions in the liquid state is determined by a specific combination of
$f_0, \tau_p$ and $\rho_a$ that provides a measure of the relative magnitude of 
the ``active'' stress with respect to the stress arising from interparticle forces; 
(iv) The phase diagram in the $T-f_0$ plane shows the complete 
disappearance of the glass phase beyond a threshold value of the activity; 
(v) The presence of activity decreases the kinetic fragility of the liquid; and (vi)
Activity leads to local clustering of the self-propelled particles induced 
by the passive particles in the glassy medium.

We study the Kob-Andersen binary mixture using 800 A-type and 200 B-type 
particles interacting via the Lennard-Jones pair potential,
\begin{equation}
 V_{ij}(r)=4 \epsilon_{ij} \[\(\frac{\sigma_{ij}}{r}\)^{12}-\(\frac{\sigma_{ij}}{r}\)^{6}\],\label{eq:potential}
\end{equation} 
where $r$ is the distance between two particles and the indices $i$, $j$ can be A or B. 
The values of $\sigma_{ij}$ and $\epsilon_{ij}$ are chosen to be: 
$\sigma_{AB}=0.8 \sigma_{AA}$, $\sigma_{BB}=0.88 \sigma_{AA}$, $\epsilon_{AB}=
1.5  \epsilon_{AA}$, $\epsilon_{BB}=0.5 \epsilon_{AA}$. We set a cut off in 
the potential at $r_{ij}=2.5 \sigma_{ij}$ and shift it accordingly. We set 
the unit of length and energy by  $\sigma_{AA}=\epsilon_{AA}=1$ and fix the 
overall density $\rho$ at $1.2$. Our MD simulation in three dimensions is 
carried out in the NVT-ensemble using a modified leap-frog integration 
scheme~\cite{Brown:84}. The ``temperature'' T for both zero and finite
activity is determined from the average kinetic energy of the
particles.

We introduce activity only through a fraction $\rho_a$ of B-type particles, 
while the A-type particles remain passive. The active B-particles are randomly 
assigned self-propulsion forces of the form ${\bf{f_0}}=f_0(k_x\hat{{\bf x}}+
k_y\hat{{\bf y}}+ k_z\hat{{\bf z}})$, where $k_x, k_y, k_z$ are $\pm 1$, 
chosen so that the net momentum of the system 
remains conserved. After a persistence time $\tau_{p}$ the directions of 
$\{\bf{f_0}\}$ are randomized by choosing a different set of $k_x, k_y, k_z$, 
while respecting momentum conservation. This discrete 8-state clock realization
of the random forces shows the same qualitative features as a continuous O(3) 
description and all the results reported here are obtained with this scheme.
 
\begin{figure}
\includegraphics[height = 0.575\linewidth]{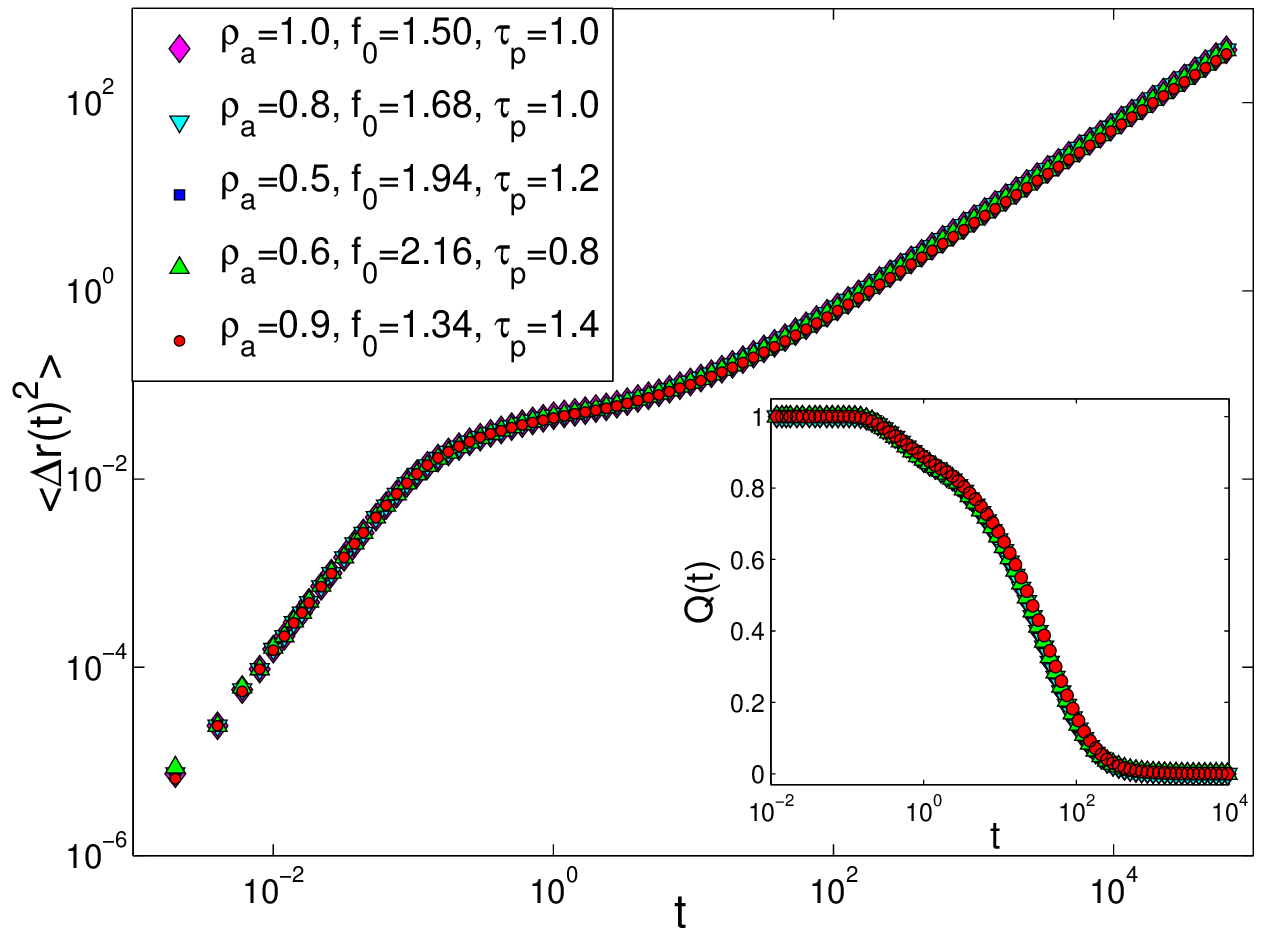}
\caption{(color online). MSD as a function of time at $T=0.5$ for five different 
sets of values of the parameters $f_0$, $\tau_p$, $\rho_a$, chosen such that 
${\rho_a}{{f_0}^2}{\tau_p}$ remains constant. The plots for different sets of parameter values collapse into
a single curve. In the inset the collapse of the two-point 
correlation function $Q(t)$ for the same sets of parameter values is shown.}
\label{collapse}
\end{figure}

\begin{figure}
\includegraphics[height = 0.575\linewidth]{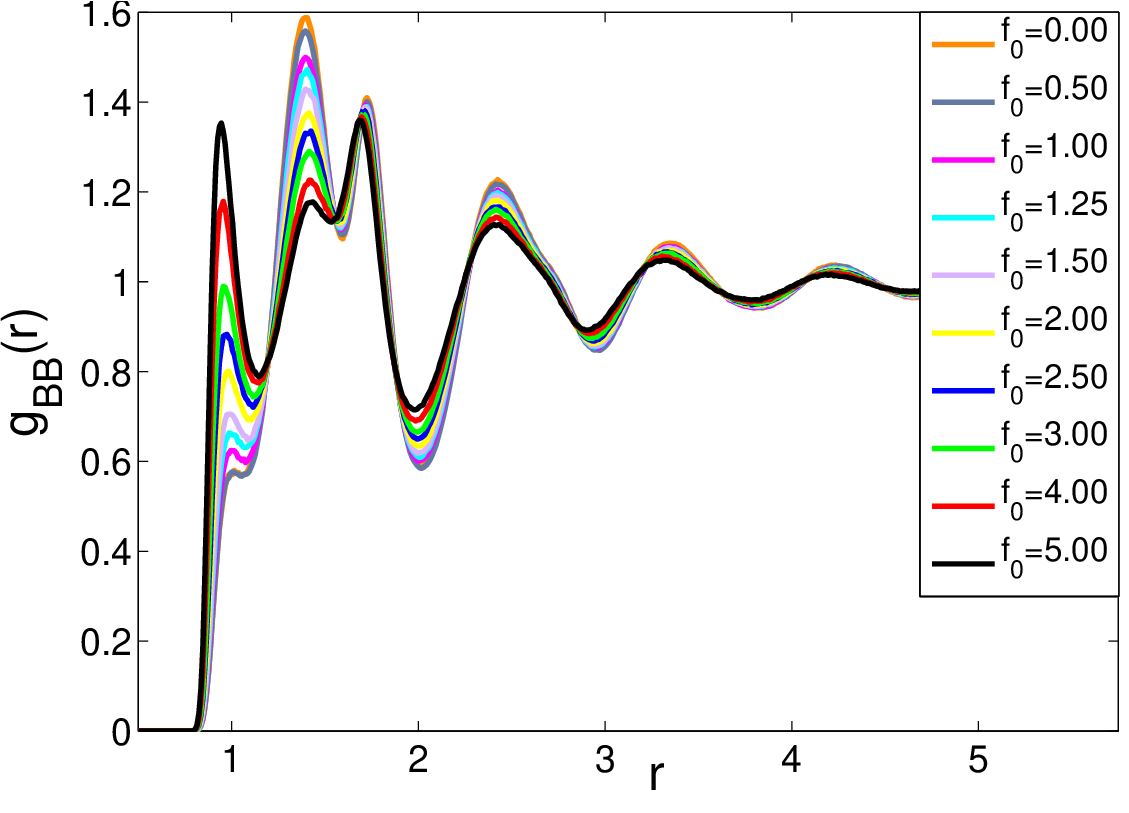}
\caption{(color online). Pair correlation function $g_{BB}(r)$ of  B-type (active) particles 
for different values of $f_0$ at $T=0.45$, $\rho_{a}=1.0$ and $\tau_{p}=4.0$. 
The appearance of a peak at a lower value of $r$ and the increase in its height with the increase of 
activity $f_0$ is indicative of activity-induced clustering.}
\label{gbbr}
\end{figure}

To study the dynamics of the system, we measure the mean-square-displacement 
(MSD) of tagged particles, $\langle |\Delta {\bf r}(t)|^2\rangle$, as a 
function of time $t$ at different temperatures. 
Simultaneously, we record the two-point correlation function, $Q(t)$, whose decay in time provides a measure of  
the dynamical slowing down in glassy systems,
\begin{equation}
Q(t) = \frac{1}{N} \sum_{i}\langle w(\mid{\bf r}_i(t_0)- {\bf r}_i(t+t_0)\mid)\rangle
\end{equation}
where,
\begin{equation}
w(r)=
\left\{
        \begin{array}{ll}
                1  & \mbox{if } r \leq a\\
                0  & \mbox{otherwise}
        \end{array}
\right.
\end{equation}
and $\langle \cdots \rangle$ represents an average over the time origin $t_0$.
Here, $N$ is the number of particles and the parameter $a$ is associated with the typical 
amplitude of vibrational motion of the particles. For all of our analysis we have used $a=0.3$.

At high enough temperatures, $T=2.0$, in the absence of activity ($f_0=0$), 
the MSD increases ballistically ($\sim t^2$) at short time scales, before 
crossing over to a diffusive ($\sim t$) regime at late times. The associated 
$Q(t)$ decays exponentially to zero, characteristic of a liquid. On decreasing 
the temperature, the MSD begins to show a small plateau at intermediate 
times which  grows as the temperature decreases further. Simultaneously, 
$Q(t)$ starts exhibiting multi-step relaxation, described by a stretched 
exponential function at long times. At very low temperatures, e.g. for $T=0.2$,
both the MSD and $Q(t)$ remain in the plateau region over the time scales 
of the simulation and do not show the late time diffusive part, indicating
that the system has entered a glassy state. The self-diffusion constant $D$, 
calculated from the long-time data for the MSD using the relation 
$\lim_{t \to \infty}\langle|\Delta {\bf r}(t)|^2\rangle= 6Dt$, shows that  
in the absence of activity, the diffusion constant decreases very rapidly 
with decreasing temperature.

\begin{figure}
\includegraphics[height = 0.62\linewidth]{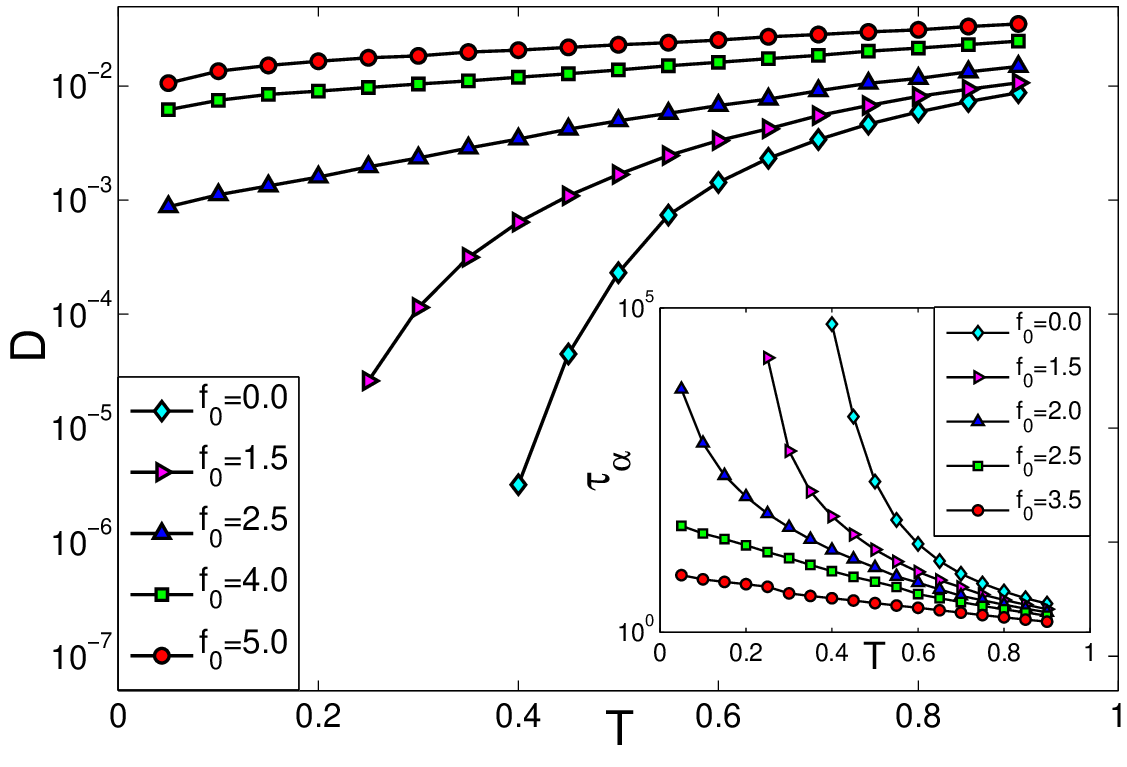}
\caption{(color online) The diffusion constant ($D$) has been plotted as a function of temperature ($T$) 
for different values of $f_{0}$ for $\tau_{P}=4.0$ and $\rho_a=1.0$. The plot shows that 
the dependence of the diffusion constant on the temperature becomes weaker with increasing $f_{0}$. The inset shows 
the $\alpha$-relaxation time as a function of temperature ($T$) for different 
$f_{0}$, illustrating a qualitative change in the behavior with increasing $f_{0}$ and a reduction in the 
(putative) glass transition temperature.}
\label{difftau}
\end{figure}

Introducing activity via self-propulsion leads to  a dramatic change in the 
dynamical behavior of the system. The behavior of the MSD and $Q(t)$, as 
illustrated in Fig.~\ref{msdqt}, clearly shows that the system fluidizes 
from the glassy state - particles escape from their cages without 
showing any sign of the intermediate plateau. At high enough activity, the 
diffusion constant shows a large enhancement and appears to be only weakly 
dependent on the temperature (see Fig.~\ref{difftau}). While activity is described by the 3 independent 
parameters  $f_0$, $\tau_p$, $\rho_a$, it is the combination  
$\zeta = {\rho_a}{{f_0}^2}{\tau_p}$ that appears to control the dynamical behavior 
in the fluidized phase.
When plotted for a fixed value of $\zeta$ 
with different choices of $f_0$, $\tau_p$, $\rho_a$, many dynamical quantities show a collapse, as shown 
in Fig.~\ref{collapse}. The quantity $\zeta$ appears naturally in a heuristic 
calculation of the effects of activity, based on a simple Langevin model 
(see the {\it Supplementary Information}). As discussed there, when this quantity is made dimensionless by dividing it by 
an appropriate combination of parameters characterizing the interparticle forces, it provides a measure of the 
relative magnitude of the active force with respect to typical interparticle forces.

\begin{figure}
\includegraphics[height = 0.60\linewidth]{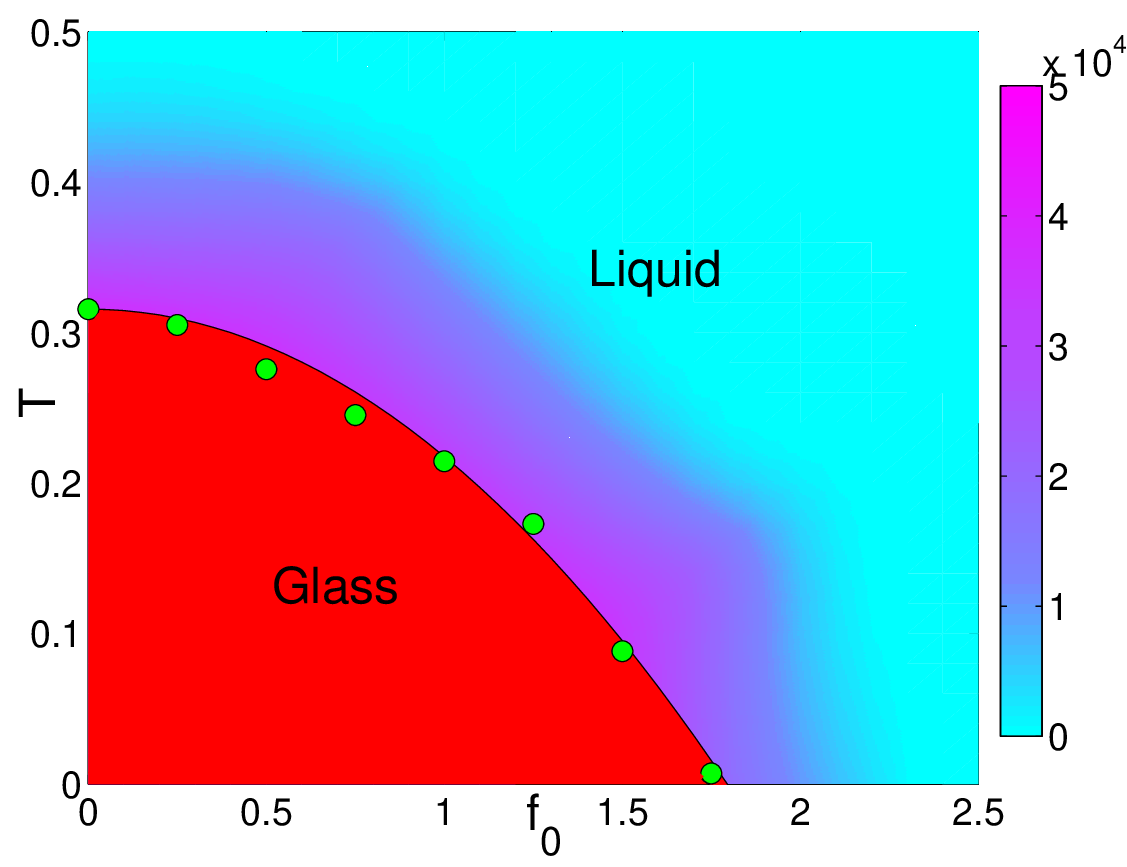}
\vspace{.1cm}
\caption{(color online). Phase diagram in the $T-f_{0}$ plane. The glass transition temperatures (green filled circles) have been obtained  by fitting the  $\alpha$-relaxation time $\tau_{\alpha}$  to extract $T_{\mbox{\tiny{VFT}}}$ for different values of $f_0$. Both dimensional considerations and heuristic arguments based on a Langevin model ({\it Supplementary Information}) suggest that 
the phase boundary has the form $T_{\mbox{\tiny{VFT}}}(f_0) =T_{\mbox{\tiny{VFT}}}(0)-A\,\rho_a \tau_p {f_0}^2$, 
where $A$ is a constant of order one. The phase boundary (thin black line) represents a fit of the data for $T_{\mbox{\tiny{VFT}}}(f_0)$ to this form.}
\label{phased}
\end{figure}

To check whether this activity induced fluidization is accompanied by a significant
change in the structure of the liquid, we have calculated
the radial distribution functions $g_{\alpha \beta}(r)$, 
defined by
\begin{equation}
g_{\alpha \beta }(r)=\frac{V}{N_{\alpha}N_{\beta}}\langle \sum \limits^{N_{\alpha}}_{i=1} \sum \limits^{N_{\beta}}_{j=1} \delta({\bf r}-{\bf r}_{i}^{\alpha}+{\bf r}_{j}^{\beta})\rangle, 
\end{equation}
where $N_\alpha$ is the number of particles of type $\alpha$ ($\alpha = A$ or $B$).
We find that  $g_{BB}(r)$ is more significantly affected by activity than 
$g_{AA}(r)$ and $g_{AB}(r)$ (Fig.~\ref{gbbr} and Fig.~S2 in
{\it Supplementary Information}), and shows more
liquid-like features as $f_0$ is increased. Also, the height of the first peak of $g_{BB}$ decreases 
and a new peak at a smaller value of $r$ develops with increasing $f_0$. 
This observation suggests that there is an activity-induced 
clustering tendency in the B-type particles, mediated by the correlations of 
the A particles. The clustering tendency is confirmed from a calculation (see the {\it Supplementary Information}) of the 
size distribution of the clusters formed by B-type particles.

The strong effect of activity on the dynamics is seen very clearly in Fig.~\ref{difftau} where the temperature
dependence of the long-time diffusion coefficient $D$ has been shown for different values of the 
self-propulsion force $f_0$. The rapid decrease in $D$ with decreasing $T$ for $f_0=0$ is replaced
by a much weaker temperature dependence for large $f_0$. At any temperature, the diffusion constant 
increases as $f_0$ is increased from zero. This behavior is qualitatively different from that reported
in Ref.~\cite{Berthier:14} where the diffusion constant was found to {\em decrease} with increasing activity
at relatively low densities (equivalent to relatively high temperatures in our system), causing a crossing of
$D$ vs. $\rho$ plots for different strengths of the activity. In our system, $D$ vs. $T$ plots for
different $f_0$ come closer to each other as $T$ is increased, but they do not cross. This behavior is 
qualitatively similar to that observed in Ref.~\cite{Ni:13}.

Activity-induced fluidization is also seen in plots of the $\alpha$-relaxation time $\tau_\alpha$, 
extracted from  the decay of $Q(t)$ ($Q(\tau_{\alpha})=1/e$), vs. $T$ for different
$f_0$ (inset of Fig.~\ref{difftau}). To extract a glass transition temperature from the data
for the $\alpha$-relaxation time, we fit $\tau_\alpha$  to the 
well-known Vogel-Fulcher-Tammann (VFT) form,
\begin{equation}
\tau_\alpha=\tau_\infty \exp \[\frac{1}{\kappa\(\frac{T}{T_{\mbox{\tiny{VFT}}}}-1\)} \]
\label{eq:VFT}
\end{equation}
where $\kappa$ is the kinetic fragility, $\tau_\infty$ is the relaxation time 
at high temperatures, and $T_{\mbox{\tiny{VFT}}}(f_0)$ is the activity
dependent (putative) glass transition temperature at which the relaxation time extrapolates
to infinity.

In the absence of activity, $T_{\mbox{\tiny{VFT}}}(0)\sim 0.3$ in reduced Lennard-Jones units. 
The values of $T_{\mbox{\tiny{VFT}}}$ for different values of $f_0$ allow 
us to  construct a phase diagram in the $T-f_0$ plane (Fig.\,\ref{phased}). This shows that even at 
low temperatures, there is a threshold activity beyond which one may exit the glassy phase into the liquid. 
The magnitude of $\tau_{\alpha}$, displayed as a color plot in Fig.~\ref{phased}, shows a sharp increase near the phase 
boundary. The region displayed in red color in Fig.~\ref{phased} represents infinite $\tau_{\alpha}$ according to the VFT form. 
In this phase diagram, the glass transition temperature, defined as the temperature at which
$\tau_\alpha$ would diverge if the VFT form continues to describe its temperature dependence, approaches
zero for a {\em finite} value of the active force $f_0$. This is qualitatively different from the behavior found
in Refs.~\cite{Berthier:14} and \cite{Ni:13} in which a (putative) glass transition was found to be present for
all finite values of the strength of the activity. This important observation tells us that some of the effects of
activity on the glass transition are sensitive to the nature of the system (whether driven by temperature or density) and
the details of the self-propulsion mechanism. This observation is consistent with general expectations derived from a  
study~\cite{kurchan} of dynamic arrest in a mean-field spin-glass model with non-thermal fluctuations.

\begin{figure}
\includegraphics[height = 0.60\linewidth]{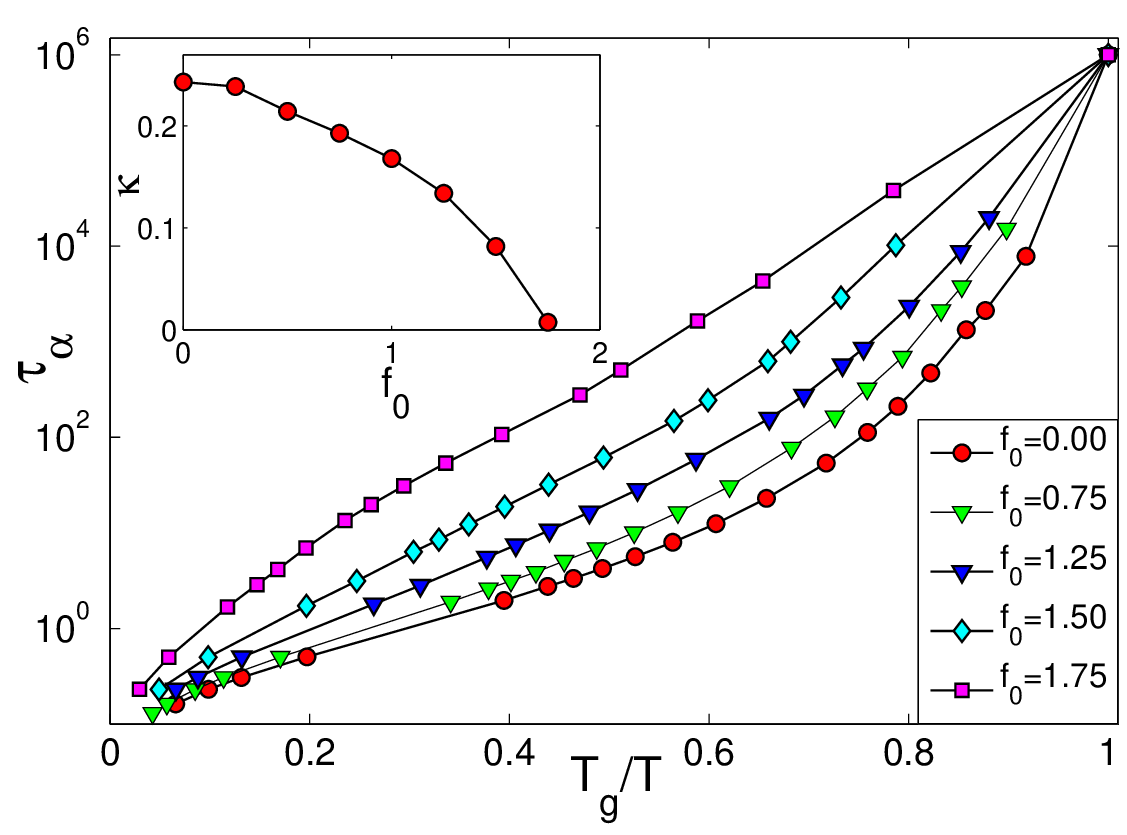}
\caption{(color online). Angell plot showing a crossover from fragile to strong behavior with increasing active 
force ($f_0$). In the inset, the kinetic fragility ($\kappa$) has been plotted as a function of $f_0$.}
\label{fragility}
\end{figure}

Figure~\ref{fragility} shows so-called ``Angell Plots''  - $\tau_\alpha$ vs $T_g/T$,  where $T_g$, the analog
of the experimentally determined glass transition temperature at which the viscosity is $10^{13}$ poise, 
is obtained from the definition $\tau_\alpha(T_g)=10^6$. The curvature of the plots decreases with increasing
$f_0$, indicating that the fragility decreases with increasing activity. The dependence of the kinetic fragility $\kappa$ 
obtained from the VFT fits on the active force $f_0$ (see the inset of Fig.~\ref{fragility}) exhibits the same trend,  
indicating that the activity makes the glass stronger. This is opposite to the behavior
of the hard-sphere system studied in Ref.~\cite{Ni:13} - the fragility was found to increase
with increasing activity in that work.

To summarize, we have demonstrated activity driven fluidization in a model glass former 
at low temperatures and a 
concomitant reduction of the glass transition temperature with increasing activity. 
This fluidization is accompanied by a crossover from caging dynamics to diffusive 
transport at long times. 
The late-time diffusion coefficient in the activity-induced 
fluid phase is weakly dependent on 
the temperature. We display a phase 
diagram in the ${T-f_{0}}$ plane that suggests that the glass transition temperature goes
to zero at a finite threshold value of $f_0$. The shape of the phase boundary has been 
rationalized from a simple calculation ({\it Supplementary Information}).
This calculation also brings forth the possibility of existence of a single 
control parameter in the form 
of $\zeta={\rho_a}{{f_0}^2}{\tau_p}$, constructed from the three parameters 
${\rho_a}$, $f_0$ and ${\tau_p}$ that
characterize the effects of activity on the dynamics of the system. In the fluid phase, 
different combinations of the parameters ${\rho_a}$, $f_0$ and ${\tau_p}$ that give a 
constant value of $\zeta$ lead to the same time dependence of
dynamical quantities such as the MSD and $Q(t)$. However, 
the same behavior is not observed below $T_{\mbox{\tiny{VFT}}}$. 
We have also observed an activity induced clustering of the self-propelled particles. 
The kinetic fragility of the glass former has been found to decrease with increase in activity.

The results obtained here are very general and should be relevant to a large variety of dense colloids where chemical 
activity can induce fluidization ~\cite{Bradley:14,Angelini:11}. This work also makes contact with many reports 
of glassy behavior and activity induced fluidization in cells and tissues. We are currently exploring variants of this 
model to explain specific features of caging and fluidization observed in the cell surface and the nucleus. 
\\
\\
We thank Saurish Chakrabarty and Akshay Bhatnagar for useful discussions. R. M. and P. J. B. acknowledges financial support from CSIR, India. C. D. acknowledges financial support from DST, India. M. R. acknowledges a grant  from the Simons Centre.

\end{document}


\title
{Active fluidization in dense glassy systems -- Supplementary Information}

\author{Rituparno Mandal}%
\email[Email: ]{rituparno@physics.iisc.ernet.in}
\iisc

\author{Pranab Jyoti Bhuyan}%
\email[Email: ]{pranab@physics.iisc.ernet.in}
\iisc

\author{Madan Rao}%
\email[Email: ]{madan@ncbs.res.in}
\rri
\ncbs

\author{Chandan Dasgupta}%
\email[Email: ]{cdgupta@physics.iisc.ernet.in}
\iisc

\maketitle
\section{Heuristic Calculation}

As described in the main text, a dense soft glass-forming liquid, modeled by the Kob-Andersen binary mixture of A and B particles, shows strong collective caging behavior at sufficiently low temperatures. Incorporating activity
or self-propulsion, $f_0$, in one set of particles, say B, induces cage breaking and fluidization. Here we try and capture some features of  this collective caging and escape, within a single-particle
description, using a Langevin dynamics for the position of the tagged-particle in an effective caging potential created by the neighboring particles. Activity enters into the Langevin equation both as a source of athermal noise and as a slower remodeling of the caging potential (Fig.\,1).

\begin{figure}[!ht]
\begin{center}
\begin{tikzpicture}[scale=0.7]
\draw[thick] (4,-0.3) rectangle (13.1,6.2);
\draw[ultra thick,<->] (9.4,3.2) -- (9.4,.9) -- (12.4,.9);
\draw[blue,thick] (9.7,2.5) .. controls (10.2,-1) and (11.2,-1) .. (11.7,2.5);

\draw[blue,thick,dashed] (11.55,1.5) .. controls (11.95,2.6) and (12.5,1.6) .. (13,1);
\draw[->,thick] (11.8,2.3) .. controls (12,2.45) and (12.5,3.4) .. (12.8,1.7);
\draw[thick,->] (10.75,0.3) -- (11.35,0.3);
\shade[inner color=red!15,outer color=red] (10.75,0.3) circle (0.25);

\node[right] at (11.4,0.55) {r};
\node[right] at (8.1,1.9) {V(r)};

\draw[thick,->] (5.6,5.7) -- (5.9,6.135);
\shade[inner color=red!15,outer color=red] (5.6,5.7) circle (0.25);

\draw[thick,->] (5.05,4.2) -- (5.05,4.75);
\shade[inner color=red!15,outer color=red] (5.05,4.2) circle (0.25);

\draw[thick,->] (6.9,5.05) -- (7.45,5.15);
\shade[inner color=red!15,outer color=red] (6.9,5.05) circle (0.25);

\draw[thick,->] (5.9,2.9) -- (5.9,2.35);
\shade[inner color=red!15,outer color=red] (5.9,2.9) circle (0.25);

\draw[thick,->] (7.25,3.75) -- (6.65,3.65);
\shade[inner color=red!15,outer color=red] (7.25,3.75) circle (0.25);

\shade[inner color=blue!15,outer color=blue] (4.85,5.7) circle (0.35);
\shade[inner color=blue!15,outer color=blue] (6.45,5.6) circle (0.35); 
\shade[inner color=blue!15,outer color=blue] (7.3,5.68) circle (0.35);

\shade[inner color=blue!15,outer color=blue] (4.62,4.9) circle (0.35);
\shade[inner color=blue!15,outer color=blue] (5.45,5.05) circle (0.35);
\shade[inner color=blue!15,outer color=blue] (6.25,4.8) circle (0.35); 
\shade[inner color=blue!15,outer color=blue] (7.85,5.1) circle (0.35);

\shade[inner color=blue!15,outer color=blue] (4.45,3.95) circle (0.35);
\shade[inner color=blue!15,outer color=blue] (5.73,4.25) circle (0.35); 
\shade[inner color=blue!15,outer color=blue] (6.64,4.18) circle (0.35); 
\shade[inner color=blue!15,outer color=blue] (7.35,4.48) circle (0.35);
\shade[inner color=blue!15,outer color=blue] (8.1,4.3) circle (0.35); 

\shade[inner color=blue!15,outer color=blue] (4.55,3.10) circle (0.35);
\shade[inner color=blue!15,outer color=blue] (5.35,3.4) circle (0.35);
\shade[inner color=blue!15,outer color=blue] (6.15,3.5) circle (0.35);
\shade[inner color=blue!15,outer color=blue] (7.85,3.4) circle (0.35);

\shade[inner color=blue!15,outer color=blue] (5.35,2.5) circle (0.35);
\shade[inner color=blue!15,outer color=blue] (6.55,2.45) circle (0.35);
\shade[inner color=blue!15,outer color=blue] (6.90,3.2) circle (0.35);
\shade[inner color=blue!15,outer color=blue] (7.57,2.7) circle (0.35);

\shade[inner color=white!15,outer color=white] (3.25,4.2) circle (0.35);
\shade[inner color=white!15,outer color=white] (3.50,5.65) circle (0.35);
\shade[inner color=white!15,outer color=white] (3.55,4.95) circle (0.35);

\shade[inner color=red!15,outer color=red] (10,5.65) circle (0.25);
\node[right] at (10.4,5.6) {B-type};
\shade[inner color=blue!15,outer color=blue] (10,4.75) circle (0.35);
\node[right] at (10.4,4.75) {A-type};

\draw[thick,->] (9.65,3.8) -- (10.45,3.8);
\node[align=left,right] at (10.4,3.5) {Self\\ Propulsion\\ Force};

\end{tikzpicture}
\end{center}

\caption{(color online). Schematic diagram (in the top left) for self-propelled dynamics and (in the right bottom) for the effective dynamics of B-type particles in the presence of activity in a cage-like environment, represented by a caging potential $V$. Activity generates an athermal noise for the B particles and remodels the caging-potential to allow for escape.}
\end{figure}
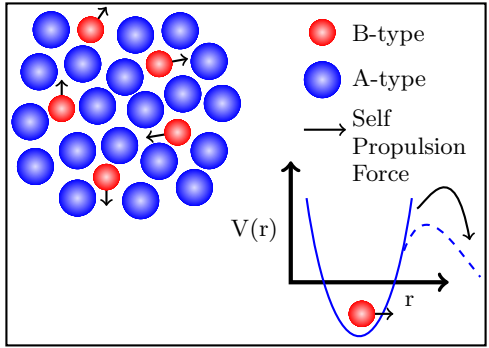

Let the fraction of passive (A) particles be $1-\rho_a$; we model the dynamics of a tagged passive particle by a Langevin equation for the displacement $x_A(t)$,
\be
m\,\ddot{x}_A=-V^{\prime} - \gamma\, \dot{x}_A+\xi(t)\, ,
\ee
where $m$ is the mass of a particle (for both A and B-type) and $\gamma$ is the effective friction coefficient. 
$V^{\prime}$ is the force derived from a caging potential whose form we comment on later.
The thermal noise 
$\xi(t)$ is taken to have zero mean and white, with a variance equal to $2\gamma k_BT$.

A fraction $\rho_a$ of particles (B) are, in addition, subject to a random active force with amplitude $f_0$. The randomness is modeled by an athermal noise $\psi(t)$, which is exponentially correlated over a time scale $\tau_p$. The Langevin equation describing its displacement is given by,
\be
m\,\ddot{x}_B=-V^{\prime}-\gamma \,\dot{x}_B+\xi(t)+f_0\,\psi(t)\, .
\ee
The statistics of the athermal noise is given by $\langle \psi(t)  \rangle=0$, $\langle \xi(t)  \psi(t^{\prime})  \rangle=0$ and 
\be
\langle \psi(t) \psi(t^{\prime}) \rangle= c \exp\left[-\frac{\vert t-t^{\prime}\vert}{\tau_p}\right] 
\ee
where $c$ is a dimensionless constant. For a fixed $\tau_p$ and $c$, this athermal noise in general does not obey the fluctuation-dissipation relation.

The caging potential $V$ (Fig.\,1) is taken to have the same form for both A and B particles, with a confining harmonic part,
$V=\frac{1}{2} k x_{\alpha}^2$ (where $\alpha=A, B$), and a barrier whose height is reduced by the active force over a time scale corresponding to the
$\alpha$-relaxation time. In what follows, we will only be concerned with the harmonic confining part of the potential,
represented by the parameter $k$.

It is straightforward to calculate the mean square displacement of the A and B particles,
\be
\langle [x_A(t)]^2\rangle =\frac{k_BT}{k}
\ee
\be
\langle [x_B(t)]^2\rangle =\frac{k_BT}{k}+\frac{f_0^2 c \tau_p}{\gamma k}\frac{1}{(1+\frac{\tau_p}{\tau_x})}\,,
\ee
where the relaxation time $\tau_x=\gamma/k$.
The particle-averaged mean square displacement may be written as,
\be
\langle x(t)^2\rangle  =  \rho_a \langle x_B(t)^2\rangle+(1-\rho_a) \langle x_A(t)^2\rangle
\ee
from which we can derive an average Lindemann factor,
\begin{equation}
\frac{\langle x^2\rangle}{a^2}=\frac{k_B T}{k a^2} + \frac{\rho_a {f_0}^2 \tau_p }{\gamma k a^2}\left[{\frac{c}{1+\frac{{\tau_p}}{{\tau_x}}}}\right]
\label{lind}
\end{equation}
Here, $a$ is the typical cage size and $k a^2$ is the potential energy scale. This gives us a heuristic phase boundary which varies with the self-propulsion force as $T_{\mbox{\tiny{VFT}}}(f_0) =T_{\mbox{\tiny{VFT}}}(0)-A\,\rho_a \tau_p {f_0}^2$; where $A$ is a proportionality constant. This simple minded calculation suggests the existence of a single control parameter, $\zeta={\rho_a}{{f_0}^2}{\tau_p}$, constructed from the three active parameters ${\rho_a}$, $f_0$ and ${\tau_p}$.  We have observed that, different combinations of the parameters ${\rho_a}$, $f_0$ and ${\tau_p}$ for a constant value of $\zeta$ leads to a collapse of the dynamical quantities like MSD and $Q(t)$ in a temperature range above the VFT temperature ($T_{\mbox{\tiny{VFT}}}$). However, we do not observe the same data collapse below $T_{\mbox{\tiny{VFT}}}$.

The dimensionless form of the parameter $\zeta$, appearing in the second term on the right-hand side of Eq.(\ref{lind}), can be written as
\begin{equation}
\frac{\zeta}{\gamma k a^2} =  \rho_a \left[ \frac{f_0}{ka}\right]^2 \frac{\tau_p}{\tau_x}.
\end{equation}
For fixed $\tau_p/\tau_x$, this quantity is proportional to the square of the ratio $f_0/(ka)$, which represents the 
relative magnitude of the active force $f_0$ with respect to the typical interatomic force $ka$. 

\begin{figure}[!ht]
\begin{center}
\includegraphics[height=.6\columnwidth]{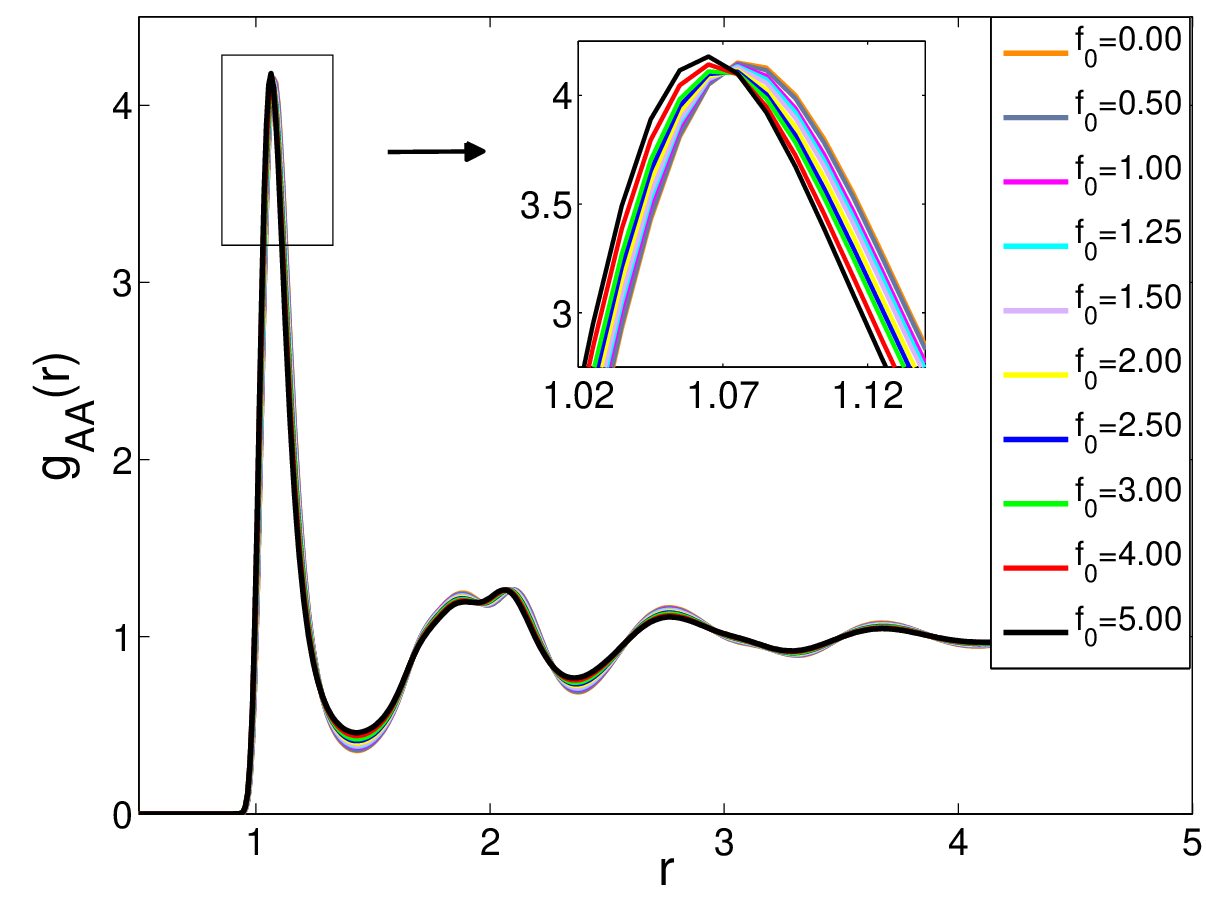}
\includegraphics[height=.58\columnwidth]{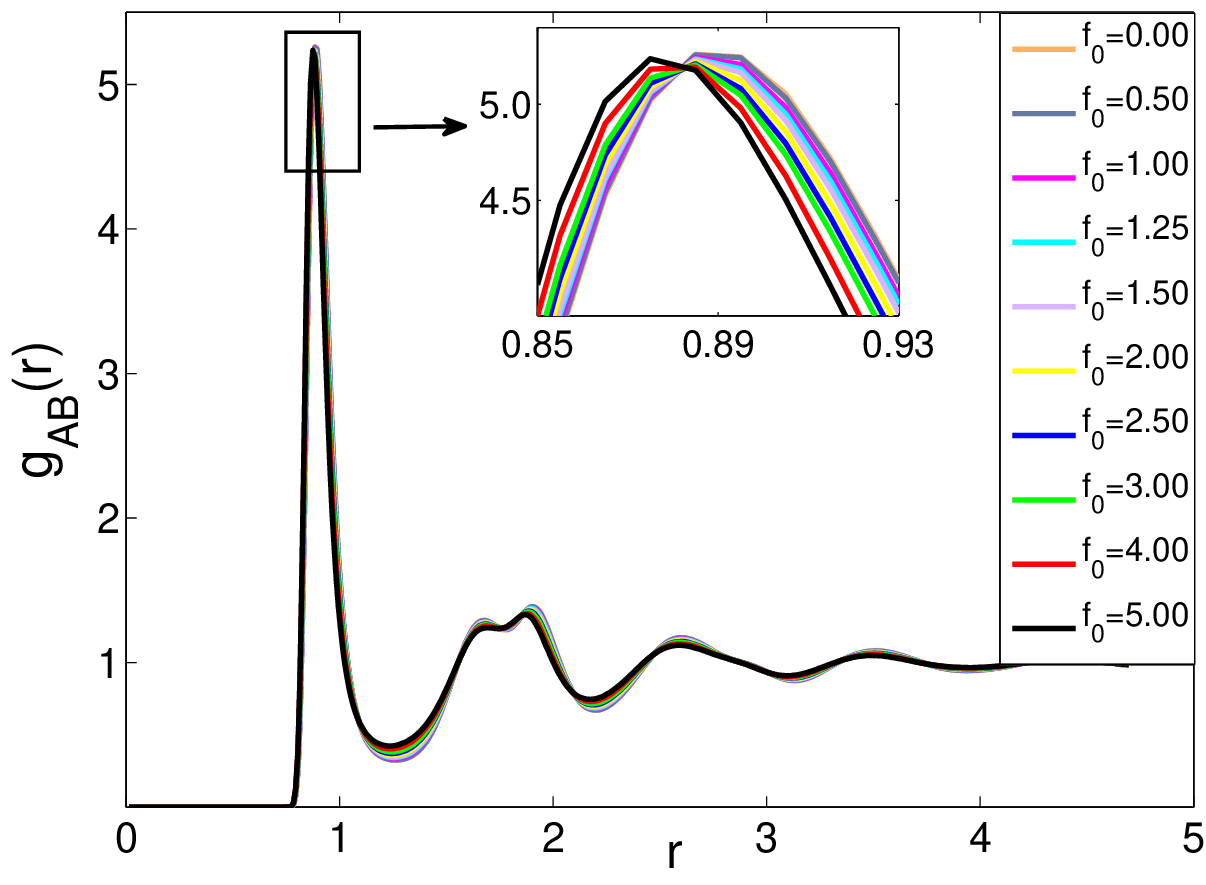}
\caption{(color online). Changes in pair correlation function (a) (top) $g_{AA}(r)$ and (b) (bottom) $g_{AB}(r)$ in the presence of different values of self-propulsion forces ($f_0$) for $T=0.45$, $\rho_{a}=1.0$ and $\tau_{P}=4.0$. Insets in the figures show the shift of the first peak of $g(r)$ to lower values of $r$ with increasing $f_0$, suggesting formation of structures at smaller scales.}
\label{gaaabr}
\end{center}
\end{figure}

\section{Pair Correlation Function}
The pair correlation function $g_{\alpha \beta}(r)$ is defined as
\begin{equation}
g_{\alpha \beta }(r)=\frac{V}{N_{\alpha}N_{\beta}}\langle \sum \limits^{N_{\alpha}}_{i=1} \sum \limits^{N_{\beta}}_{j=1} \delta({\bf r}-{\bf r}_{i}^{\alpha}+{\bf r}_{j}^{\beta})\rangle, 
\end{equation}
where $N_\alpha$ is the number of particles of type $\alpha$ ($\alpha = A$ or $B$).

We find that, $g_{AA}(r)$ and $g_{AB}(r)$ are less affected by the presence of activity compared to $g_{BB}(r)$, which shows a significant change with increasing activity. This suggests that presence of stochastic activity does not affect the local structure of the passive or A-type particles in any significant manner, whereas the B-type particles are affected much more strongly. 

\section{Evidence of activity induced clustering}
To understand the decrease of the first peak height and the simultaneous development of a new peak at a smaller value of $r$ in $g_{BB}(r)$ , we analyze the cluster size distribution for the B-type particles. For this purpose, we define two B-particles to be connected if their interparticle distance is less than a cutoff $r_0$, chosen to be slightly larger than the value of $r$ at the new peak in $g_{BB}(r)$ that appears with increasing activity (we have 
taken $r_0$=1.0 for the results shown below).

\begin{figure}[!ht]
\begin{center}
\includegraphics[height=.66\columnwidth]{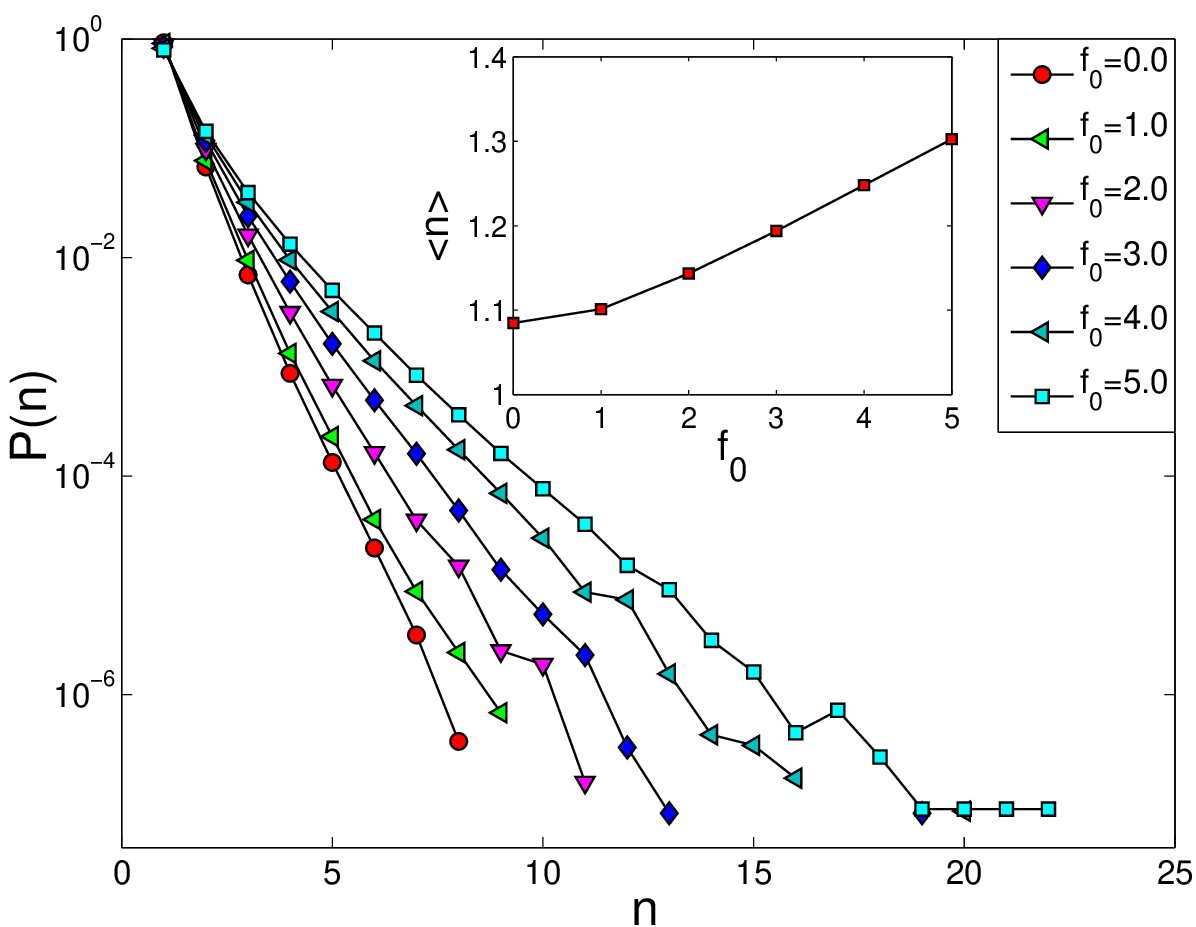}
\caption{(color online). Cluster size distribution function $P(n)$, where $n$ is the size of a cluster of the B-type particles, for different $f_0$ at $T=0.45$, $\rho_a=1.0$ and $\tau_p=4.0$. With increasing activity ($f_0$), the distribution broadens, signifying the clustering tendency of B-type particles. In the inset the average cluster size has been plotted as a function of $f_0$.}
\label{cluster}
\end{center}
\end{figure}